\documentstyle[twocolumn,aps,epsf,floats]{revtex}
\begin{document}
\draft
\title{Quasienergy Spectroscopy of Excitons}
\author
{Kristinn Johnsen and Antti-Pekka Jauho}
\address{
Mikroelektronik Centret, Technical University of Denmark, Bldg
345east\\
DK-2800 Lyngby, Denmark\\
\medskip
\date{Draft: \today}
\parbox{14cm}{\rm
We theoretically study nonlinear optics of excitons
under intense THz irradiation. In particular, the linear
near infrared absorption and resonantly enhanced
nonlinear sideband generation are described.
We predict a rich structure in the spectra which can be interpreted
in terms of the quasienergy spectrum of the exciton, via a remarkably
transparent expression for the susceptibility, and
show that the effects of strongly avoided quasienergy crossings
manifest themselves directly, both in the absorption and transmitted
sidebands.
\smallskip\\}
}
\maketitle
\narrowtext
Excitons, strongly correlated pairs of electrons and holes,
dominate the band-edge optics of semiconductors:
their bound states are observed
optically as resonances in the absorption spectrum in
the gap of the semiconductor and in the continuum
as enhanced absorption\cite{exciton}.
The interband susceptibility describes the linear
optical properties of the semiconductor and can be expressed
in terms of the well-known equation\cite{Elliot},
\begin{equation}
\chi(\omega ) = -\sum_n\frac{|\psi_n(\vec{r}=\vec{0})|^2}{
\hbar\omega -E_n +i0^+},
\label{eq1}
\end{equation}
where $\psi_n(\vec{r})$ are the Wannier wave functions of the exciton with 
spectrum $E_n$.
This compact formula is extremely useful
since it relates the spectrum of the exciton directly to the resonances 
occurring in the absorption spectrum, which is proportional to the imaginary
part of (\ref{eq1}).

The development of coherent sources of intense electromagnetic radiation
in the THz regime has opened up a new exciting area of semiconductor
physics\cite{FELS}: e.g.\ two color spectroscopy of excitons
in the dynamical Franz-Keldysh (DFK) regime\cite{NOR98}.
One color has weak intensity in the near infrared (NIR) regime 
inducing interband transitions, creating the excitons. 
The other color, which has high intensity, is in the
far infrared regime (FIR), which corresponds to transitions between
the internal \mbox{degrees} of freedom of the created excitons. 
In this regime the field strength, $\vec{E}_{\mathrm{FIR}}$, and the
frequency, $\Omega$, of the FIR are such that both strong field
effects (dc Franz-Keldysh (FK) effect\cite{FKE}) and 
multiphoton processes (MP)\cite{MP} are important. This regime
is quantified by\cite{yacoby}
\begin{equation}
\gamma = \frac{e^2{E}_{\mathrm{FIR}}^2}{4 m_r\hbar\Omega^3},
\end{equation}
which is the ratio between the ponderomotive energy and the energy of
a FIR photon. Here $m_r$ is the reduced mass of the exciton and
$e$ is the electronic charge. The regime of FK corresponds to
$\gamma\gg 1$, of MP to $\gamma\ll 1$ while $\gamma\sim 1$ corresponds
to the regime we consider presently, the DFK regime.

Study of the response
of the weak NIR probe as a function of the two frequencies and the
intensity of the FIR beam yields considerable insight into the 
nature of excitons and the fundamental electro-optical processes
occurring in semiconductors. Experiments have been reported
on effects of the FIR on luminescence\cite{b1}, 
absorption\cite{NOR98}, and resonantly enhanced nonlinear mixing of
the fundamental NIR with the FIR\cite{NOR97}, even in the presence of 
strong magnetic fields\cite{junprl}.
The free particle properties of such experiments have been 
studied intensively\cite{yacoby,REB85,JAU96,JOH98,XU98}, while
excitonic effects on absorption have only been reported
recently\cite{MEIER95,NOR98,JOH97}. 

The goal of this letter is to report a generalization
of the text-book result (\ref{eq1}) to the new dynamical
regime $\gamma\sim 1$.  We find that 
the  macroscopic polarization can be expressed as
\begin{equation}
\vec{P}(t) = d^2\vec{E}_{\mathrm{NIR}}
\sum_n\chi_n(\omega )e^{i(\omega + n\Omega )t}
\label{macP}
\end{equation}
where
\begin{equation}
\chi_n(\omega ) = -\sum_{\alpha n'}
\frac{\phi^*_{\alpha n'+n}\phi_{\alpha n'}}{
\hbar\omega-\tilde\epsilon_\alpha-n'\hbar\Omega+i0^+}.
\label{chin}
\end{equation}
Here $\omega$ is the frequency of the NIR, 
$d$ is the interband dipole matrix element, assumed to be constant, and
$\phi_{\alpha n}$ are the temporal Fourier components of the
Floquet states $\phi_\alpha(\vec{r}=\vec{0},t)$
of the exciton, with {\em quasienergy} 
$\tilde\epsilon_\alpha$, \cite{Floquet}.
A Floquet state is the temporal analogue to a Bloch state in a
spatially periodic potential. 
The spectral decomposition
of the polarization, (\ref{macP}), shows that the linear absorption
of the NIR is proportional to Im${\chi_0(\omega )}$, while
$I_n(\omega) = |\chi_n(\omega )|^2$ is proportional to the
intensity of photonic sidebands which radiate at frequency
$\omega + n\Omega$. 
Eq.\ (\ref{chin}) thus relates the nonequilibrium optical properties
directly to the excitonic quasienergy spectrum.

We first outline the derivation of (\ref{macP}) and
then present a numerical 
study of a quantum well (QW)
exciton, which illustrates the applicability of the theory. 
We show that {\em avoided crossings} in the quasienergy
spectrum lead to two clear experimental consequences: (i)
the sideband intensities $I_n$ have a resonant behaviour and (ii)
the $1s$ resonance in optical absorption displays a strong
 Autler-Townes splitting\cite{Autler55}.

In the following we describe the semiconductor via a simple
spinless effective mass two-band model. The Hamiltonian is given by
\begin{equation}
H = \sum_{\vec k,\mu\in\{c,v\}}
\epsilon_\mu[\hbar\vec k+e\vec A(t)]c_{\mu\vec k}^\dagger c_{\mu\vec k}
+ H_i,
\end{equation}
where $\mu$ labels conduction- or valence-bands, and $H_i$
describes the  Coulomb interaction via the potential $v(|\vec k|)$.
Using nonequilibrium Green function methods one can show that
the  retarded susceptibility obeys the integral equation\cite{integr},
\begin{eqnarray} 
\label{eq:integr}
\chi^r (\vec k;t,t') = \bar\chi^r (\vec k;t,t') +
\quad\quad\quad\quad\quad\quad\quad\quad\quad\quad\quad\\
 \int \frac{d^n\vec k'}{(2\pi)^n}\int dt''\, \bar\chi^r (\vec k;t,t'')
v(|\vec k-\vec k'|)\chi^r (\vec k';t'',t'),
\nonumber
\end{eqnarray}
where $\bar\chi^r (\vec k;t,t')$ is the 
freeparticle susceptibility\cite{footnote}.

Eq.\ (\ref{eq:integr}) can be transformed into an infinite-dimensional
matrix equation and solved numerically by introducing a cut-off
frequency\cite{NOR98,JOH97}.
This method has successfully described some effects in the linear 
absorption\cite{NOR98},
but has failed to describe the nonlinear mixing adequately.
The representation introduced here does not introduce cutoffs,
is numerically much faster, and clarifies the physics involved.
In the rotating wave approximation with respect to the NIR probe,
the irreducible interband susceptibility for an undoped 
semiconductor obeys\cite{JOH98}
\begin{equation}
\biggl\{
i\hbar\frac{\partial}{\partial t} -\epsilon[\hbar\vec k+e\vec A(t)]
\biggr\}\bar\chi^r (\vec k;t,t') = \delta (t -t'),
\label{eq:chi0dyson}
\end{equation}
where 
$\epsilon(\hbar\vec{k}) = \epsilon_c(\hbar\vec{k})
-\epsilon_v(\hbar\vec{k})$ and $e=|e|$.
The uniform FIR field
is described by the vector potential
$\vec A(t) = -\vec E_{\text{FIR}}\sin (\Omega t) / \Omega$.
With (\ref{eq:chi0dyson}) we transform (\ref{eq:integr}) to
\begin{eqnarray}
\label{eq:chidiff}
\biggl\{
i\hbar\frac{\partial}{\partial t}
-\epsilon[\frac{\hbar}{i}\vec\nabla_{\vec r}+e\vec A(t)]
+\frac{e^2}{4\pi\kappa r}
\biggr\}\chi^r (\vec r;t,t')  \quad\quad\\
\quad\quad\quad\quad\quad\quad\quad\quad\quad
\quad\quad\quad\quad\quad\quad
= \delta (\vec r)\delta (t -t'),
\nonumber
\end{eqnarray}
where $\kappa$ is the effective
dielectric constant in the semiconductor. To proceed we first
consider the solutions to the homogeneous part of (\ref{eq:chidiff}),
\begin{equation}
\biggl\{
\frac{[\frac{\hbar}{i}\vec\nabla_{\vec r}+e\vec A(t)]^2}{2m_r} +\epsilon_g
-\frac{e^2}{4\pi\kappa r}
\biggr\}\Psi(\vec r,t) = i\hbar\frac{\partial \Psi(\vec r,t)}{\partial t}.
\label{eq:scrho}
\end{equation}
This is a Schr\"{o}dinger equation
for a hydrogen-like particle,
the exciton, in the presence of the intense FIR field. 
When the FIR field is absent this
equation reduces to the Wannier equation\cite{exciton}.
We have thus arrived at a generalized Wannier equation
with a time periodic Hamiltonian,
$H(t) = H(t+T)$, where $T = 2\pi/\Omega$. Such a discrete time
symmetry gives rise to the temporal analog of the Bloch wave functions, 
the Floquet states\cite{Floquet}:
$
\psi_\alpha(\vec r,t) = e^{-i\tilde\epsilon_\alpha t/\hbar}
\phi_\alpha(\vec r,t),
$
where $\tilde\epsilon_\alpha$ are the quasienergies and 
$\phi_\alpha(\vec r,t) = \phi_\alpha(\vec r,t+T)$. The Floquet
states can be viewed as the stationary states of the periodically
 driven system.
They form complete sets of solutions to the 
Schr\"{o}dinger equation,
which is to say that any wave function obeying (\ref{eq:scrho})
is of the form 
$\Psi(\vec r,t) = \sum_\alpha c_\alpha e^{-i\tilde\epsilon_\alpha t/\hbar}
\phi_\alpha(\vec r,t),
$
where $c_\alpha$ are $c$-numbers. Furthermore, at equal
times the states fulfill the closure relation
$
\sum_\alpha 
\phi_\alpha^*(\vec r,t)\phi_\alpha(\vec r',t)
 = \delta (\vec r - \vec r')
$.
Using these properties,
we expand the susceptibility in terms of the
Floquet states and their quasienergies and find
\begin{equation}
\chi^r (\vec r;t,t') =
\frac{\theta (t-t')}{i\hbar}\sum_\alpha
e^{-i\tilde\epsilon_\alpha (t-t')/\hbar}
\phi_\alpha^*(\vec{0},t')
\phi_\alpha(\vec{r},t).
\end{equation}
We next set $\vec{r}=\vec{0}$, define
$\phi_\alpha(t) = \phi_\alpha (\vec{0},t)$ and 
expand the states as $\phi_\alpha(t) = \sum_n\phi_{\alpha n}e^{-in\Omega t}$.
The macroscopic polarization is then readily expressed as (\ref{macP}) and
(\ref{chin}). This concludes the outline of our derivation.

\begin{figure}[t!]
\begin{center}
\epsfxsize=7.5cm\epsfbox{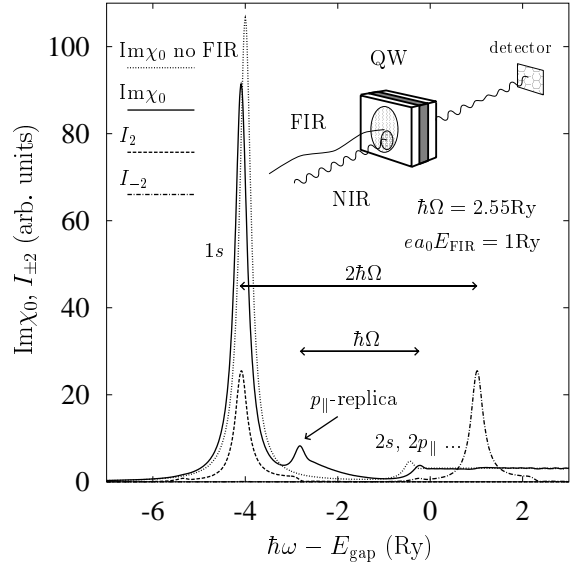}
\end{center}
\caption{
The main effects due to intense FIR irradiation, colinear
with weak NIR field.
Linear optical absorption Im$\chi_0$:
(i) ac-Stark red shift of the $1s$ resonance,
(ii) DFK blue shift of the $2s$ resonance and the band edge,
(iii) suppression of oscillator strength, and (iv)
emergence of a photon replica of the $p_\|$ state.
Nonlinear mixing $I_{\pm2}$: Sideband emission at $\omega \pm n\Omega$.
The inset illustrates the experimental geometry.
}
\label{figIntro}
\end{figure}

We next apply the present theory to the nonlinear optical
properties of a quantum well exciton.
Specifically we choose 
$ea_0E_{\mathrm{FIR}} = 1E_{\mathrm{Ry}}$,
where $E_{\mathrm{Ry}} = \hbar^2/(2m_ra_0^2)$ is the effective
Rydberg energy of the medium, and $a_0 = \hbar^2\kappa/(e^2m_r)$ 
is the effective Bohr radius
of the exciton. As an example, for InGaAs $E_{\mathrm{Ry}}\sim2-3$meV and
$a_0 \sim 200$\AA\ which leads to $E_{\mathrm{FIR}}\sim 10^5$V/m,
well within the range of free electron lasers\cite{FELS}. 
The FIR is linearly polarized and the field oscillates in the
plane of the QW.
We sweep the THz frequency such that it probes the various 
internal resonances of the exciton and study the absorption and the
nonlinear sideband generation.

A two dimensional exciton in
equilibrium has the bound state 
spectrum $E_n = E_{\mathrm{gap}}-E_{\mathrm{Ry}}/(n-0.5)^2$, $n>0$. 
The $n=1$ state is a nondegenerate $s$-state, 
while $n\neq 1$ are degenerate
containing also $p$-states, etc.~\cite{b6}. 
With linear polarization of the FIR the doubly degenerate
$p$-states may be decomposed into
$p_\perp$ and $p_\|$, where the $p_\|$ has the same 
spatial symmetry as the field. Only $p_\|$ contributes to the dynamics.
We include both $s$ and $p_\|$ states
in our calculations since both are 
physically important as they couple strongly
in the presence of the THz field\cite{footnote2}.
The Floquet states and their
quasienergies are determined numerically. 
In the figures, we have introduced a
phenomenological damping of the resonances in (\ref{chin}).
In Fig.~\ref{figIntro} we
show the results of our calculation
for fixed $\hbar\Omega = 2.55E_{\mathrm{Ry}}$ 
as a function of NIR frequency,
illustrating the basic effects due to the THz field. 
Considering the absorption,
the THz frequency is below the $1s\rightarrow 2p_\|$ equilibrium 
transition frequency, $\hbar\omega^0_{12}\approx 3.56E_{\mathrm{Ry}}$.
The red shift of the $1s$ resonance,
away from its equilibrium position,
is due to
the ac-Stark effect\cite{stark}, which is pronounced
even though the frequency is considerably detuned from the
$1s\rightarrow 2p_\|$ transition. 
The effect is maximal in the $\gamma\sim 1$ regime.
The $2s$ resonance is blue-shifted with respect to the equilibrium
possition.
The band edge is blue shifted as well.
This is due to the DFK effect which shifts all main features by the 
ponderomotive energy\cite{yacoby,REB85,JAU96}. 
The oscillator strength of the $1s$ resonance is suppressed.
These effects have been observed in quantum well excitons\cite{NOR98}.
Furthermore, a new resonance appears in the absorption
spectrum: a single photon replica of the dark $p_\|$ state
which under irradiation becomes optically active. The tail to the blue
from the resonance is due
to replicas of the $p_\|$ symmetric states in the continuum.
The sideband intensity for $I_{\pm 2}$ is shown as well.
In reflection symmetric systems only sidebands with $n$ even appear:
the optical properties are invariant 
under the transformation 
$\vec E_{\mathrm{FIR}}\rightarrow -\vec E_{\mathrm{FIR}}$,
and hence the ``effective'' frequency is
$2\Omega$. Breaking the reflection symmetry introduces odd 
sidebands as well. The sideband generation is at a maximum if
either $\omega$ or $\omega-n\Omega$, $n>0$, is tuned to the
$1s$ resonance. 
We find that $I_{n}(\omega ) = I_{-n}(\omega + n\Omega)$.
However, in an experiment using a multiple quantum well
sample the radiation due to $I_{-n}$ will 
tend to be re-absorbed, its maximum
being tuned to the main absorption resonance 
while $I_n$ has its maximum 
$2n\Omega$ away from it. This behavior of the
sideband generation has been observed\cite{NOR97}.

\begin{figure}[t!]
\begin{center}
\epsfxsize=7.5cm\epsfbox{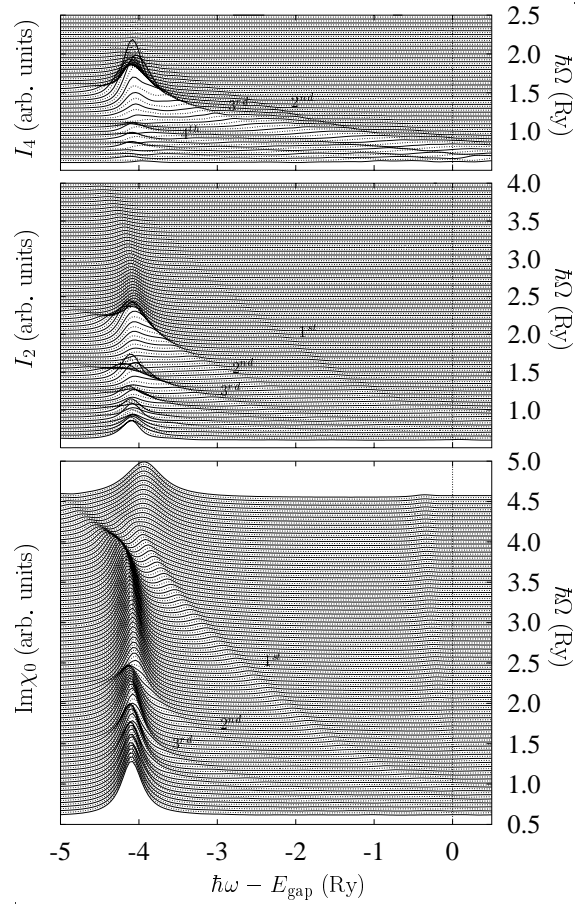}
\end{center}
\caption{
The effect of the THz frequency at fixed field strength.
From below; the absorption Im$\chi_0$,
$2$-, and $4$-photon sideband generation.
The THz frequency is indicated on the right vertical axis.
Here several avoided crossings in the quasienergy spectra
can be distinguished, as splittings in the absorption, and as resonant
enhancement in the sideband generation.
The order of the processes involved is indicated.
$I_4$ is multiplied by $3$
compared to $I_2$.
}
\label{figSweep}
\end{figure}
In Fig.\ \ref{figSweep} we show the spectra for a constant field
strength with a THz frequency sweep 
$\hbar\Omega = (0.5\cdots 5)E_{\mathrm{Ry}}$.
In Im$\chi_0$ one observes how photon replicas 
traverse from $\hbar\omega\sim E_2$ in a fan as the THz
frequency is increased. 
The fanblades have been tagged
with the order in the THz frequency involved, which is roughly 
proportional to the inverse slope of the blade.
In view of (\ref{chin}),
when the replicas reach the main resonance a strongly avoided 
crossing in the quasienergy spectrum results,
which is directly visible in the spectra. 
For the first order process a Autler-Townes splitting results. 
The sidebands, in the upper two panels, 
show clear evidence of the fan shape.
In this case however, the underlying avoided crossing
in the quasienergy spectrum results in
resonantly enhanced sideband generation. Generation
of sideband $I_n$ calls for optical processes of order $|n|+1$
or higher and thus the predominant order
determines if a strongly avoided crossing results in
a resonantly enhanced sideband.

\begin{figure}
\begin{center}
\epsfxsize=7.5cm\epsfbox{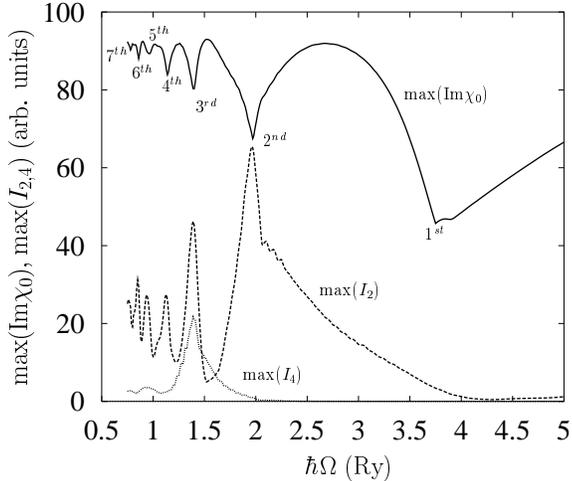}
\end{center}
\caption{The maximum absorption and sideband generation
as a function of the FIR frequency illustrating the
relation between the absorption suppression and
the resonantly enhanced sideband generation.}
\label{figMax}
\end{figure}

The oscillator strength reflects the resonant conditions. 
In Fig. \ref{figMax} we compare the oscillator strength to the 
maximum of the generated sidebands at each frequency. 
We remark in passing that the $n=2$ sideband was recently detected
for $\hbar\Omega \sim 4E_{\mathrm{Ry}}$\cite{NOR97}, which 
in view of Fig.\ \ref{figMax} suggests that
the predicted results should be
detectable with present technology.
Resonant suppression of the 
oscillator strength occurs when the sideband generation is enhanced. 
The resonance conditions
are met by the quasienergies according to (\ref{chin}).
The range in which resonant enhancement occurs is
determined by the coupling of the internal exciton levels.
The enhancement mainly involves the $1s\rightarrow 2p_\|$ transition
for odd processes, 
while it is mainly the $1s\rightarrow 2s$ transition which
is responsible for even resonances.
The first 7 resonances are indicated in Fig.\ \ref{figMax}.

In summary, we have studied the nonlinear optics of excitons
subject to intense THz radiation. We have shown analytically that 
observed resonances in the sideband generation
may be viewed as manifestations of strongly avoided crossings in
the quasienergy spectrum of the exciton.
Our theory is consistent with experiment, where available, and
it leads to a number of predictions for the outcome of new
experiments. An example is the strong correlation between
the oscillator strength suppression in absorption and the 
associated resonant enhancement in sideband generation, respectively.

We would like to acknowledge our colleagues,
S.J.~Allen, 
B.~Birnir,
B.~ Y.-K.~Hu,
A.~Ignatov,
J.~Kono,
W.~Langbein,
K.~Nordstrom,
M.~Sherwin,
and
M.~Wagner
for sharing details of their experiments and many
enlightening discussions. 


\begin{thebibliography}{8}
\bibitem{footnote}
The present letter is concerned with undoped semiconductors, where the low 
density limit is appropriate. It is possible to formulate the theory for 
screened interaction; this would lead to a renormalized form of the results 
presented here.
\bibitem{footnote2}
Note that the exciton state created due to a single photon
interband transition is always a $s$-state. Coupling
to $d$-, $f$-states, and so on, are thus of third order or higher
and due to degeneracy dominating
equivalent $s$- to $p$-state transitions are
always present. Therefore only $s$- and $p$-states are
included in the calculations.
\end{thebibliography}
\end{document}